\documentstyle[aps,prl,twocolumn]{revtex}

\begin{document}
   
\title{Decoherence, wave function collapses 
and non-ordinary statistical mechanics}
\author{Mauro Bologna$^{1}$, Paolo Grigolini$^{1,2,3}$, 
Marco G. Pala$^4$, Luigi Palatella$^{2}$}
\address{$^{1}$Center for Nonlinear Science, University of North Texas,
   P.O. Box 311427, Denton, Texas 76203-1427 }
\address{$^{2}$Dipartimento di Fisica dell'Universit\`a di Pisa and
INFM, via Buonarroti 2, 56126 Pisa, Italy}
\address{$^{3}$Istituto di Biofisica CNR, Area della Ricerca di Pisa,
Via Alfieri 1, San Cataldo 56010 Ghezzano-Pisa, Italy}
\address{{$^{4}$} Dipartimento di Ingegneria dell'Informazione 
dell'Universit\`a di Pisa, via
Diotisalvi 2, 56122 Pisa, Italy}
\date{\today}
\maketitle

   \begin{abstract}
 We consider a toy model of pointer interacting with a $1/2$-spin system,
 whose $\sigma_{x}$ variable is \emph{measured} by the
environment, according to the prescription of decoherence theory. 
If the environment measuring the variable $\sigma_{x}$ yields
ordinary statistical mechanics, the pointer sensitive to the $1/2$-spin
 system undergoes the same, exponential, relaxation regardless of whether
real collapses  or an entanglement with the environment, mimicking the effect
 of real collapses, occur. In the case of non-ordinary statistical mechanics
 the occurrence of real collapses make the pointer still relax exponentially
 in time, while the equivalent picture in terms of reduced density matrix
 generates an inverse power law relaxation.  
   \end{abstract}
\vspace{0.5 cm}
Pacs: 03.65.Ta, 03.67.-a, 05.20.-y, 05.30.-d \\ 

Decoherence theory was born in 1970 with the seminal work of Zeh 
\cite{seminalwork} and grew up with the work of Zurek and others over
 the following decades.  It is now regarded to be a theory so robust 
as to make Tegmark and Wheeler \cite{scientificamerican} claim 
that it renders obsolete
 the hypothesis of wave function collapses made by
 the founding fathers of quantum mechanics. The main purpose of this paper
 is to prove that this claim is correct only in the case when decoherence 
is caused by interactions compatible with ordinary statistical mechanics.
 If we move from the condition of exponential relaxation, which 
is the key property of ordinary statistical mechanics, to the condition
 of inverse power law relaxation, the statistical equivalence between 
wave-function collapses and decoherence is lost.
To show this basic property,  let us consider the following toy Hamiltonian
\begin{equation}
H_{T} = G \Sigma_{z} \sigma_{x} +H_{\sigma },
\end{equation}
where
\begin{equation}
\label{thehamiltonianofthesystemofinterest}
H_{\sigma} = -\frac{V}{\sqrt{2}} \sigma_{z} + g \sigma_{x} \eta + H_{B} .
\end{equation}
We have a $1/2$-spin system, characterized by the Pauli matrix 
$\bf \Sigma$, called \emph{pointer}, interacting with a $1/2$-spin system, 
characterized by the Pauli matrix $\bf \sigma$, and called \emph{system of 
interest}. 
The system of interest undergoes an interaction with a bath, 
through a variable $\eta$,
 driven by the Hamiltonian $H_{B}$. The density matrices of the pointer 
and of the
 system of interest  are called $\rho_{\Sigma}$ and $\rho_{\sigma}$, 
respectively. 
The former is obtained from a contraction over the degrees of freedom 
of the system
 of interest, and of its bath as well. 
The latter requires a contraction over the
 pointer degrees of freedom as well as on the bath of the system of interest. 
 This bath is 
assumed to be much faster than the pointer and, as a consequence, the 
time evolution of $\sigma_{x}$ is virtually independent of the pointer 
dynamics.

First of all, we show that in the special case where the correlation function
 of the fluctuation $\eta$ is exponential, the two pictures, wave-function
 collapses and decoherence, yield the same statistical result. 
This supports the point of view  of the advocates of decoherence. 
Then, we create a condition of anomalous statistical mechanics, 
by modulating the Hamiltonian $H_{\sigma}$ in such a way as to create
 a significant departure from ordinary exponential relaxation. 
In this case, we show that the two perspectives yield quite different results.
  In a sense, the decoherence theory is not contradicted, 
in so far as the pointer 
density matrix becomes diagonal in the basis set of the pointer eigenstates. 
However, this happens via either an exponential relaxation, if the system of 
interest undergoes real collapses,  or through an inverse power law decay, 
if no real collapses occur. 

The case  of wave-function collapse in the Markovian case  
was already discussed in an earlier publication\cite{marcoluigi}, 
where it was proved that the bath of the system of interest, 
making measurements with frequency $1/\tau$, where 
$\tau = (1/2) [\hbar^{2}/(g^{2} \langle \eta^{2} \rangle \tau_{\eta})]$
\cite{note2} and $\tau_{\eta}$ is the 
correlation time of $\eta$,
results in a sequence of symbols such as 
+,+,+,-,-,-,...
This is so because the observable $\sigma_{x}$  has the 
eigenstates $|+ \rangle_{x}$ and $|- \rangle_{x}$, with the eigenvalues 
$1$ and $-1$, respectively. The inphasing term $-V \sigma_{z}/\sqrt{2}$ 
forces the system of interest to stay in a superposition 
of both states, with the form
\begin{equation}
\label{inphasing}
|\psi(t)\rangle = |+\rangle_{x} \cos (\omega t + \phi) - i  
|-\rangle_{x} \sin(\omega t + \phi).
\label{genericform}
\end{equation}
If the first measurement is done at $t = 0$, and the system 
of interest makes an instantaneous collapse into $|+\rangle_{x}$,  
for instance, then the subsequent time evolution of the system 
of interest is given by Eq.(\ref{inphasing}) with $\phi = 0$.  
We set the condition  $\tau \ll 1/ \omega$ (overdamped condition), 
with $\omega \equiv V/ \sqrt{2} \hbar$.  This means that the next measurement, 
occurring at $t = \tau$, will probably make the system collapse into 
$|+\rangle_{x}$ again.  In fact, the overdamped condition makes it possible 
for us to evaluate the probability of collapse into $|-\rangle_{x}$ by Taylor 
series expansion of $\sin(\omega t)$, and this yields, for the probability of 
the system to collapse into $|-\rangle_{x}$, 
the value $(\omega \tau)^{2} \ll 1$.
This is a small, but non-vanishing, quantity. 
Thus, sooner or latter the system 
of interest will collapse into $|-\rangle_{x}$. 
If it does, it will keep collapsing 
into $|-\rangle_{x}$ several more times. The distribution of waiting times 
in either of these two states is expressed by \cite{marcoluigi}
\begin{equation}
\psi(t) = \frac{\gamma}{2} \exp(-\frac{\gamma}{2} t), \quad {\rm with} \quad
\gamma \equiv 2 \omega^{2} \tau. 
\label{distributiondensity}
\end{equation}
Note that this picture in terms of wave function collapses is 
compatible with the Lindblad equation \cite{lindblad} that reads
\begin{equation}
\label{liblandequation}
\frac{d}{dt} \rho_{\sigma} = 
L_{\sigma} \rho_{\sigma} (t) = - i \omega [ \sigma_{z}, \rho_{\sigma} ] 
- \frac{1}{2 \tau} [ \sigma_{x},[ \sigma_{x}, \rho_{\sigma} ] ].
\end{equation}

It is important to notice that
\begin{equation}
\Phi_{\sigma}(t) = \langle\sigma_{x}(0) \sigma_{x}(t)\rangle
\end{equation}
can be evaluated, with no assumption of wave-function collapses, 
from Eq. (\ref{liblandequation}). In the specific case
when the overdamped condition applies, the operator $L_{\sigma}$ 
has virtually only two eigenstates, namely $0$ and $\gamma$, 
and it is easy to prove that
\begin{equation}
\label{overdampedcase}
\Phi_{\sigma}(t) = \exp(-\gamma t).
\end{equation}
On the other hand the renewal theory \cite{geisel} allows us to establish a 
connection between the waiting time distribution $\psi(t)$ and the 
correlation function $\Phi_{\xi}(t)$ of the 
sequence of eigenvalues $\xi(t)$ of 
$\sigma_{x}$, generated by the wave-function collapses.
We have
\begin{equation}
\label{renewaltheory}
\Phi_{\xi}(t) = \frac{1}{\langle t \rangle} 
\int_{0}^{t}(t-t^{'})\psi_{r}(t^{'}) {\rm d}t^{'}.
\end{equation}
The mean value $\langle t \rangle$ is the mean waiting time determined by 
$\psi_{r}(t)$. The Laplace transform of $\psi_{r}$, 
$\hat \psi_{r}(s)$, 
is related to the Laplace transform of $\psi(t)$, $\hat \psi(s)$, 
by \cite{zumofenklafter} 
\begin{equation}\label{Laplacerelation}
\hat \psi_{r}(s)=2 \frac{\hat \psi(s)}{1+\hat \psi(s)}.
\end{equation}
This is so because the renewal theory implies a random choice of sign 
at the end of the laminar region, namely $\psi_{r}(t)$, while $\psi(t)$ 
refers to the alternated sign condition.
 It is straightforward to show that 
$\psi_{r}(t) = \gamma \exp (-\gamma t)$, $\langle t \rangle = 1/\gamma$ and 
$\Phi_{\xi}(t) = \exp (-\gamma t) = \Phi_{\sigma}(t)$, the latter 
equality being another manifestation of the equivalence between 
decoherence theory and wave-function collapses.
 A further significant aspect 
of this equivalence has been pointed out in Ref.\cite{marcoluigi}. 
These authors discussed a physical condition equivalent to the 
Hamiltonian of Eq. (2), and studied the transition process from an 
initial out-of equilibrium condition with 
$\langle +|_{x}\rho_{\sigma}(0)|+\rangle_{x}$ 
being the only non-vanishing component 
of the reduced density matrix, to the final equilibrium, with the two 
states $|+\rangle_{x}$ and $|-\rangle_{x}$
 equally populated. They found that the von 
Neumann entropy increases with a rate identical to the 
Kolmogorov-Sinai (KS) entropy.
Since the KS entropy
 is a trajectory property and the von Neumann entropy is a density
 property, this equivalence supports the claim\cite{scientificamerican} 
that decoherence theory makes the real existence 
of wave-function collapses unnecessary.

Let us establish an even deeper support for this claim, 
by studying the pointer dynamics.  
Let us suppose that the initial pointer state is
\begin{equation}
|U\rangle_{\Sigma} = \alpha |+ \rangle_{z} + \beta |+ \rangle_{z},
\end{equation}
with $|+ \rangle_{z}$ and $|- \rangle_{z}$ denoting the 
eigenstates of $\Sigma_{z}$, with eigenvalues $1$ and $-1$, respectively.
The reduced density matrix of the pointer, $\rho_{\Sigma}(t)$, is the sum of 
a diagonal, $D$,  and of an off-diagonal part, $Q$, as follows:
\begin{equation}
\rho_{\Sigma}(t) = D(t) + Q(t),
\end{equation}
with
\begin{equation}
D(t) = |\alpha|^{2} |+ \rangle_{z} \langle+|_{z} 
+ |\beta|^{2}|-\rangle_{z}\langle - |_{z}
\end{equation}
and
\begin{equation}
Q(t) = \beta \alpha^{*}|-\rangle_{z}\langle +|_{z} R(t) + \alpha \beta^{*} 
|+ \rangle_{z} \langle -|_{z} R^{*}(t).
\end{equation}
The pointer relaxation $R(t)$ is given by
\begin{equation}
\label{pointerrelaxation}
R(t) =\left \langle T 
\exp \left ( \frac{2iG}{\hbar} x(t) \right ) \right \rangle,
\end{equation}
where $T$ denotes time ordering and
\begin{equation}\label{xdefinition}
x(t) \equiv \int_{0}^{t} \sigma_{x}(t')dt',
\end{equation}
with
\begin{equation}
\sigma_{x}(t) \equiv \exp \left ( \frac{i}{\hbar}H_{\sigma}t \right )
\sigma_{x} \exp \left (-\frac{i}{\hbar}H_{\sigma}t \right ).
\end{equation}
If the interaction between system of interest and its bath
 causes the wave-function collapses of the system of interest,
then the time evolution of $\sigma_{x}(t)$ becomes a classical 
fluctuation $\xi(t)$, holding either the value  $\xi = +1$ or
 $\xi = -1$ and $x(t)$ becomes a classical random walk
 trajectory driven by the diffusion equation
\begin{equation}
\label{diffusionequation}
\frac{\partial}{\partial t} p(x,t) = 
\frac{\langle\xi^{2}\rangle}{\gamma} 
\frac{\partial^{2}}{\partial x^{2}} p(x,t).
\end{equation}
In this case with standard Gaussian integration 
Eq.(\ref{pointerrelaxation}) yields
\begin{equation}
\label{advocatesright}
R(t) = \exp \left ( -\frac{2G^{2}}{\hbar^{2} \omega^{2} \tau}t \right ).
\end{equation}
On the other hand, if we do a quantum calculation with no wave-function
 collapse assumption,  for the only purpose of determining
the time evolution of the reduced density matrix $\rho_{\Sigma}(t)$,
we obtain the Lindblad form \cite{lindblad}
\begin{equation}
\label{markovapproximation}
\frac{\partial}{\partial t} \rho_{\Sigma} = 2 \frac{G^{2}}{\hbar^{2}} 
\hat \Phi_{\sigma}(0) [ \Sigma_{z} \rho_{\Sigma} (t) \Sigma_{z} 
-  \rho_{\Sigma}(t) ],
\end{equation}
where $\hat \Phi_{\sigma}(s)$ denotes the Laplace 
transform of the correlation function of Eq.(\ref{overdampedcase})

By evaluating the Laplace transform of this correlation function at
 $s = 0$ and plugging it in Eq.(\ref{markovapproximation}), we obtain for 
$R(t)$ the same expression as that of Eq.(\ref{advocatesright}), thereby 
giving an even stronger support to the claims of the advocates of decoherence.
The pointer variable $\Sigma_{z}$ is measured by the $1/2$-spin system. 
Consequently, the pointer density matrix becomes diagonal in the basis 
set of the two eigenstates of this variable. Moreover, the corresponding 
relaxation process is the same exponential relaxation,  regardless of 
whether the $1/2$-spin system has been made to collapse or not, 
by its thermal bath.

We prove now that the condition of complete equivalence between 
wave-function collapses and decoherence theory is lost
 in the case of anomalous statistics.
To simulate a condition of anomalous statistical mechanics, as 
done recently by Beck\cite{beck}, we modulate the exponential 
rate with a very slow fluctuation that has the effect of 
changing the exponential relaxation into an inverse power law. 
This means that the Hamiltonian $H_{\sigma}$ of 
Eq.(\ref{thehamiltonianofthesystemofinterest}) must 
be replaced by the Hamiltonian $H_{\sigma, \zeta}$, defined by
\begin{equation}
\label{thehamiltonianofthesystemofinterest2}
H_{\sigma, \zeta} = -\frac{V}{\sqrt{2}} \sigma_{z} + 
g(\zeta) \sigma_{x} \eta + H_{B}(\eta,\zeta) + H_{\zeta},
\end{equation}
where we introduce an interaction with a slow modulating environment, 
driven by the Hamiltonian $H_{\zeta}$, so as to create a 
distribution of $\gamma$, rather than a single $\gamma$, 
corresponding to ordinary statistical mechanics, and to 
the exponential functions 
of Eqs.(\ref{overdampedcase}) and (\ref{renewaltheory}).
We set, for the time scales
$\tau_{\eta}$, $\tau_{\sigma}$, 
$\tau_{\zeta}$ and $\tau_{\Sigma}$, of the fluctuation $\eta$, 
of the $1/2$-spin system, of the modulating variable $\zeta$, 
and of the pointer, respectively, the important condition
\begin{equation}
\tau_{\eta} \ll \tau_{\sigma} \ll \tau_{\zeta} \ll \tau_{\Sigma}.
\label{importantcondition}
\end{equation}

To make a quantum mechanical prediction in general it 
is necessary to establish the form of the correlation function 
$\Phi_{\sigma}(t)$. Thus, to 
make a prediction without using the wave-function collapse assumption, 
we must establish the form that  $\Phi_{\sigma}(t)$ gets in 
this anomalous case. To assign to this correlation function a 
simple analytical form, we choose for the equilibrium 
distribution of $\gamma$ the following form\cite{beck}
\begin{equation}
p_{eq}(\gamma) = 
\frac{T^{\mu-1}}{\Gamma(\mu-1)} \gamma^{\mu-2} \exp(-\gamma T).
\label{luigi'sproposal}
\end{equation}
Actually, we assume the ergodic condition to apply, and consequently 
the correlation function $\Phi_{\sigma}(t)$ can also be evaluated 
through an average in time, as follows:
\begin{equation}
\Phi_{\sigma}(t) = \lim_{L \rightarrow \infty} 
\frac{1}{L} \int\limits_0^{L} \exp [-\gamma(t')t]dt',
\label{luigi'sintegration}
\end{equation}
thereby implying that the time spent in a physical condition 
characterized by a given value of $\gamma$ is proportional 
to $p_{eq}(\gamma)/(\gamma \langle 1/\gamma \rangle)$. 
Thus, we evaluate the integral 
of Eq.(\ref{luigi'sintegration})  by using the
following prescription
\begin{equation}
\Phi_{\sigma}(t) = \left \langle \frac{1}{\gamma} \right \rangle^{-1}
\frac{T^{\mu-1}}{\Gamma(\mu-1)}
\int\limits_{0}^{\infty} {\rm d}\gamma \frac{d\gamma}{\gamma} 
e^{-\gamma T} \gamma^{\mu-2} e^{-\gamma t}.
\end{equation}
At this stage it is straightforward to prove that the 
correlation function $\Phi_{\sigma}(t)$ gets the inverse power law form:
\begin{equation}
\label{luigikeyproposal}
 \Phi_{\sigma}(t) = \left ( \frac{T}{t+T} \right )^{\beta} 
{\rm with} \:\: \beta = \mu -2.
\end{equation}

We focus our attention on a condition as far  as
 possible from that of ordinary statistical mechanics. 
This condition is naturally given by
\begin{equation}
\label{anomalousstatisticalmechanics}
0 < \beta < 1,
\end{equation}
which makes the correlation function $\Phi_{\sigma}(t)$ 
non integrable. This has deep consequences. Notice that the
physical condition that we are considering corresponds to a 
slow modulation of the two-state operator
enforced by the overdamped condition. Thus, it is is easy 
to prove that in this case 
Eq.(\ref{markovapproximation}) has to be replaced by the more general form:
\begin{equation}
\label{nonmarkovapproximation}
\frac{d}{dt} \rho_{\Sigma} =  k \int_{0}^{t} dt'  \Phi_{\sigma}(t-t') 
[ \Sigma_{z} \rho_{\Sigma} (t') \Sigma_{z} -  \rho_{\Sigma}(t') ],
\end{equation}
with $k = 2 G^{2}/\hbar^{2}$.
This non-Markovian equation yields for the relaxation $R(t)$ 
of the pointer  the following equation of motion:
\begin{equation}
\frac{dR}{dt} = - k \int_{0}^{t} \Phi_{\sigma}(t-t') R(t') dt'. 
\end{equation}
The Laplace transform of $R(t)$, $\hat R(s)$ is connected to 
the Laplace transform of $\Phi_{\sigma}(t)$, $\hat\Phi_{\sigma}(s)$, by
\begin{equation}
\hat R(s) =  \frac{1}{s + k \, \hat \Phi_{\sigma}(s)}.
\label{laplace}
\end{equation}
Using a method based on fractional derivatives, detailed in 
Ref.\cite{juri}, we show that in the limiting case $s \rightarrow 0$,
\begin{equation}
\hat \Phi_{\sigma}(s) = 
T \left [ \frac{\Gamma(1-\beta)}{(sT)^{1-\beta}} - 1 \right ].
\end{equation}
We remind the reader that we are considering 
the condition of Eq. (\ref{luigikeyproposal}).
 In this case $\hat \Phi_{\sigma}(s)$ diverges for $s \rightarrow 0$. 
This means that 
$s$ in Eq. (\ref{laplace}) can be neglected and Eq.(\ref{laplace}) yields
\begin{equation}
\hat R(s) = \frac {1} { k T} \frac{(sT)^{1-\beta}}{\Gamma(1-\beta)}.
\end{equation}
Using the Tauberian theorem\cite{weiss}, we arrive at the main conclusion 
that in the time asymptotic limit ($t \rightarrow \infty$)
\begin{equation}
R(t) \propto \frac{1}{t^{2-\beta}}.
\label{nocollapse}
\end{equation}

What about the case when wave-function collapses occur? 
If they do occur, the important property to evaluate becomes the waiting time 
distribution $\psi_{r}(t)$. According to the modulation prescription, 
this important function turns out to be given by
\begin{equation}
\label{effectivewaiting}
\psi_{r}(t) = 
\int\limits_{0}^{\infty} {\rm d}{\gamma} \frac{T^{\mu-1}}{\Gamma (\mu-1)} 
\gamma^{\mu-2} \exp(-\gamma T) \gamma \exp(-\gamma t) .
\label{waitingtimedistributionCalculus}
\end{equation}
Notice that it is straightforward to prove 
that this waiting time distribution reads
\begin{equation}
\label{simplerform}
\psi_{r}(t) =  (\mu-1) \frac{T^{\mu-1}}{(t+T)^{\mu}},
\label{waitingtimedistribution}
\end{equation}
which is related to the correlation function of Eq.(\ref{luigikeyproposal}) by 
the renewal theory prescription of Eq.(\ref{renewaltheory}). This means that 
the correlation function $\Phi_{\xi}(t)$ of the fluctuations 
$\xi(t)$ created by the wave-function collapses is identical 
to the quantum mechanical correlation, evaluated without using 
the collapse assumption. This is apparently a very reassuring 
condition for the advocates of the decoherence theory. In fact, 
it generates the impression that even in this case the direct 
use of the wave-function collapses is unnecessary. 

It is not so. The fact that the existence of trajectories 
enforces the use of $\psi(t)$ rather than that of 
$\Phi_{\sigma}(t) = \Phi_{\xi}(t)$, 
has impressive consequences, in this case.
The important relation of Eq.(\ref{pointerrelaxation}), 
if wave-function collapses occur, 
becomes identical to the characteristic function
$\langle \exp(ikx)\rangle_{t}$ of the 
corresponding diffusion process, determined by the 
the waiting time distribution
$\psi(t)$. In this case, the waiting time distribution has the form 
of Eq. (\ref{waitingtimedistribution}). We remind the reader that we 
are referring ourselves to the case where the condition $2 <\mu < 3$ applies. 
This means that the second moment of the waiting time distribution $\psi(t)$ 
is divergent. We cannot use the ordinary central limit theorem to evaluate 
this characteristic function. We can, however, use the generalized central 
limit theorem\cite{gnedenko} and, following Ref.\cite{mario,allegro,note},
we obtain
\begin{equation}
R(t) = \exp(-b|K|^{\mu-1}t),
\label{mario}
\end{equation}
where $K = 2G/\hbar$ and $b = T^{\mu-2} \sin [ \pi(\mu-2)/2 ] \Gamma(3-\mu)$. 
In conclusion, in the case of anomalous statistical mechanics, the pointer 
relaxation turns out to be exponential or an inverse power law, according 
to whether wave -function collapses do or do not occur. 

This is a striking result since it implies that a real experiment 
might make it possible to assess if wave-function collapses occur or 
not. If the experiment assessed the power law nature of the pointer 
relaxation, it would confirm the validity of quantum mechanics in a 
condition where the equivalence between decoherence theory and 
wave-function collapses is broken. If, on the contrary, the 
experiment yielded for the pointer an experimental relaxation, this 
would support the existence of real wave-function collapses, and 
these wave-function collapses, in turn, could not be entirely 
attributed to an environmental effect.
This is an issue of fundamental importance that might be resolved by 
means of real experiments, since there are currently realistic projects, 
for instance the cantilever experiment of Ref. \cite{cantilever} 
with pointers (the cantilever) sensitive 
to the fluctuations of also a single spin. 

Financial support from ARO, 
through Grant DAAD19-02-0037, is gratefully acknowledged.


\begin{references}
\bibitem{note2} As in \cite{marcoluigi}, 
the heuristic arguments here adopted are based on the 
assumption that the time distance between one measurement and the 
next is fixed, with no consequence on the theoretical prediction.
\bibitem{note} Note that the authors of Ref. \cite{allegro} proved that 
a slowly decaying oscillatory contribution to the pointer relaxation 
exists, due to the important fact that the diffusion process 
is characterized by a finite propagation front. However, 
this term becomes irrelevant with increasing the time 
scale separation between  system of interest and pointer.
\end{references}
\end{document}